\documentclass[sigconf,authorversion,nonacm]{acmart}

\usepackage{balance} 
\usepackage{adjustbox}
\usepackage{booktabs}
\usepackage{multirow}
\usepackage{bm}

\setcopyright{none}

\title[AAMAS-2024 Formatting Instructions]{Formatting Instructions for the 23rd International Conference on Autonomous Agents and Multiagent Systems}

\author{Benjamin Patrick Evans}
\affiliation{
  \institution{JP Morgan AI Research}
  \city{London}
  \country{United Kingdom}}
\email{benjamin.x.evans@jpmorgan.com}

\author{Sumitra Ganesh}
\affiliation{
  \institution{JP Morgan AI Research}
  \city{New York}
  \country{USA}}
\email{sumitra.ganesh@jpmorgan.com}

\usepackage{graphicx} 

\usepackage{booktabs}
\usepackage{subcaption}
\usepackage{caption}
\usepackage{subcaption}

\usepackage{amsmath}
\usepackage{amsfonts}

\DeclareMathOperator{\DKL}{\text{D}_\text{KL}}

\usepackage[capitalise]{cleveref}

\title{Learning and Calibrating Heterogeneous Bounded Rational Market Behaviour with Multi-Agent Reinforcement Learning}

\begin{abstract}
Agent-based models (ABMs) have shown promise for modelling various real world phenomena incompatible with traditional equilibrium analysis. However, a critical concern is the manual definition of behavioural rules in ABMs. Recent developments in multi-agent reinforcement learning (MARL) offer a way to address this issue from an optimisation perspective, where agents strive to maximise their utility, eliminating the need for manual rule specification. This learning-focused approach aligns with established economic and financial models through the use of rational utility-maximising agents. However, this representation departs from the fundamental motivation for ABMs: that realistic dynamics emerging from bounded rationality and agent heterogeneity can be modelled. To resolve this apparent disparity between the two approaches, we propose a novel technique for representing heterogeneous processing-constrained agents within a MARL framework. The proposed approach treats agents as constrained optimisers with varying degrees of strategic skills, permitting departure from strict utility maximisation. Behaviour is learnt through repeated simulations with policy gradients to adjust action likelihoods. To allow efficient computation, we use parameterised shared policy learning with distributions of agent skill levels. Shared policy learning avoids the need for agents to learn individual policies yet still enables a spectrum of bounded rational behaviours. We validate our model's effectiveness using real-world data on a range of canonical $n$-agent settings, demonstrating significantly improved predictive capability.
\end{abstract}


\begin{document}

\pagestyle{fancy}
\fancyhead{}

\maketitle

\section{Introduction}

Agent-based models (ABM) have achieved significant success in various domains, including business, epidemiology \cite{hoertel2020stochastic}, economics, and finance \cite{geanakoplos2012getting, axtell2022agent}. However, despite these achievements, criticisms persist within different communities \cite{an2021challenges}, notably in economics \cite{leombruni2005economists, Kling_2018, turrell2016agent}, primarily due to concerns regarding the decision-making rules in these systems. Frequently, these rules are manually specified heuristics, placing substantial reliance on the modeller's judgement as the simulation results and validity depend upon the specific behavioural rules utilised  \cite{turrell2016agent, osoba2020modeling}.  On the other hand, adaptive agents that optimise a utility function find greater acceptance across disciplines, as the agents' behaviour is automatically derived in a principled manner.

Hence, introducing adaptive and learning agents into ABM, while allowing for heterogeneity and bounded rationality, could alleviate these concerns and improve the realism of the models. Recent progress in reinforcement learning (RL) helps to bring this closer to reality \cite{dehkordi2023using, an2023modeling, tilbury2023reinforcement}. However, some significant challenges must be addressed before this can happen. In this work, we address the following essential question: \textit{How can we learn heterogeneous bounded rational behaviours in an ABM?}

To address this question, we introduce a novel multi-agent RL (MARL) approach where agents exhibit skill heterogeneity \cite{rogers2009heterogeneous}, constrained by their strategic processing costs. In the limit, where these processing costs $\to 0$, perfectly rational mutually consistent equilibrium can be approximated. With uniform prior beliefs and homogenous processing costs among agents, we can approximate quantal response type equilibrium. With processing costs $\to \infty$, agents act based on their prior beliefs, e.g., driven by heuristics \cite{gigerenzer2011heuristic} or biases \cite{enke2023cognitive}. However, more generally, we can model a wide range of realistic behaviour with heterogenous agent bounds. This framework aims to enhance the simulation of complex social systems, which differs from many MARL methodologies focused on learning optimal behaviours. Instead, the work aligns with the literature on ABMs, focusing on understanding the resulting realistic dynamics emerging from human decision-making.

\textbf{Contributions:} We propose an approach to effectively learn heterogeneous agent skill levels (exhibiting diverse deviations from prior beliefs) within a MARL framework. We demonstrate the efficacy of the approach across several fundamental multi-agent economic environments. The proposed approach offers substantially enhanced accuracy in predicting human decisions (along with the subsequent dynamics) in controlled experiments compared to current state-of-the-art RL approaches and other equilibrium benchmarks. To improve efficiency, we utilise agent supertypes \cite{vadori2020calibration} and shared policy learning to learn diverse bounded rational behaviours. Heterogeneity is introduced by varying strategic processing costs of agents, measured through regularised divergences from their prior beliefs. This approach is general, expanding upon a new MARL framework for modelling complex systems, \textit{Phantom} \cite{ardon2023phantom}.

\section{Related Work}\label{secBackground} 
While MARL algorithms have made significant advances in approximating equilibria within complicated environments \cite{perolat2022mastering}, a limitation is that most prevailing methodologies assume agents to be perfectly rational. This assumption is often overly stringent when simulating complex social systems \cite{arifovic2018heterogeneous}, and may miss crucial real-world dynamics \cite{bouchaud2008economics}. To broaden the applicability of these approaches for modelling complex systems, our objective is to extend MARL frameworks to account for agent heterogeneity and bounded strategic abilities. In doing so, we make the connection with ABMs while automating some of the difficult modelling design decisions (e.g., determining the behavioural rules of the agents) using RL. Recent comprehensive examinations covering RL techniques in ABM are featured in \cite{zhang2021synergistic, turgut2023framework, prasanna2019overview}, emphasizing the usefulness of learning agent behaviours.

Behavioural economics has developed more realistic models of decision-making than the traditional \textit{homo economicus} perfectly rational representative agent \cite{levitt2008homo}. Instead, these models operate under the framework of bounded rationality. One prominent approach for relaxing the strict perfectly rational Nash equilibrium (NE) assumption and incorporating bounds in reasoning is the Quantal Response Equilibrium (QRE) \cite{mckelvey1995quantal, mckelvey1998quantal}, which allows for deviations from optimal responses and the possibility of erroneous play. Such approaches generally feature consistent (and common) beliefs among agents, for example, by having the same processing costs across the population (mutual consistency). However, the need to consider agent heterogeneity has been stressed due to the ability to "bring about new outcomes not foreseeable from analysis of the homogeneous dynamics" \cite{mckelvey2000effects}. Given that populations inherently encompass a spectrum of behaviours and beliefs, a "representative" agent often proves insufficient \cite{golman2011quantal}. Recognizing the importance of strategic diversity, various extensions have been developed to accommodate a range of agent behaviours \cite{rogers2009heterogeneous, rampal2023heterogeneous}, relaxing the mutual rationality and mutual consistency assumptions \cite{chong2016generalized, Evans2023}. Relaxing mutual consistency is beneficial for multi-agent settings, allowing for a population characterized by varying levels of strategic \textit{boundedness} \cite{10.5555/3491440.3491498,latek2009bounded,patrick2023bounded}. However, the computability of such game-theoretic models is often limited to relatively simple domains, prompting the use of approximation methods (such as MARL) within more complicated environments \cite{zheng2022ai}.

Despite the achievements in behavioural game theory and RL, a gap persists in combining these methods with ABMs, e.g. for learning the decision-making rules. While RL approaches often prioritize convergence towards rational equilibrium, this focus contrasts with the primary objective of the ABM community: understanding the  properties emerging from heterogeneous boundedly rational agents. To the best of our knowledge, the seminal work on capturing realistic bounded behaviours in MARL is that of \cite{mu2022modeling}, based on rational inattention (RI). The authors of \cite{mu2022modeling} underline that the existing body of work "fails to address the modeling of bounded rationality for more accurate [MARL] simulations," proposing an innovative framework to address this limitation. Our work differs in some important ways. Specifically, we remove the difficult processing cost estimation required in \cite{mu2022modeling} (as discussed in \cref{secDiscussion}) and allow for arbitrary prior beliefs for encoding behavioural biases. Additionally, we introduce agent skill heterogeneity, learnt through regularised policies, and propose an approach for efficiently calibrating these policies to real world dynamics, all features yet to be considered.  

\section{Proposed Approach}\label{secMethod}

We introduce a novel MARL approach to effectively model a diverse range of bounded rational behaviours (or varying \textit{skill}). This approach proves valuable for calibrating ABMs to real-world systems through learning regularised policies and, under limiting cases, establishing links to various existing equilibrium solution concepts.

Our general formulation is as follows. We focus on $N$-agent systems, where each agent $i \in N$ seeks to maximise their reward (or \textit{utility}) function $U_i$ by taking actions from their action space $a \in A$. Importantly, these agents may not act perfectly rationally. The system is characterised by a state space $S$, and agents possess a (potentially partial) observation of the current state $s_i$ along with prior beliefs about their potential actions $q_i$ (a prior probability distribution over the action space). The behaviour of each agent is governed by their policy $\pi_i$, which is a mapping from states to a distribution over actions. Agents act based on their policy $a_i \sim \pi_i$, receiving reward $U_i(a_i, s_i)$. Again, these actions may not be perfectly rational and instead may be \textit{satisficing} \cite{simon1979rational,caplin2011search}.

\subsection{Components}

\subsubsection{Reward}\label{secFormulation}

In standard RL, agents aim to learn an optimal policy that maximises their expected cumulative discounted reward $U$:
\begin{equation}\label{eqOriginal}
\pi_i^*(a|s_i) = \max_{\pi_i}\mathbb{E}_{\pi_i}\left[\sum_{t=0}^\infty\gamma^t U(a_t|s_{i,t}) \right]
\end{equation}

However, often, in real world systems, the agents we seek to model are boundedly rational, driven by prior beliefs and a limited amount of processing power to improve upon these priors. To address these considerations, we incorporate generic limitations in agents' reasoning abilities by reformulating the maximisation problem into a constrained one: \footnote{We maintain exponential discounting here. However, hyperbolic discounting might provide a better alignment with human decision-making \cite{rubinstein2003economics}, although current RL approaches have not yet exhibited significant differences between the two \cite{fedus2019hyperbolic}.}

\begin{equation}
\begin{aligned}
& \pi_i^\lambda(a|s_i) = \max_{\pi}\mathbb{E}_{\pi_i}\left[\sum_{t=0}^\infty\gamma^t U(a_t | s_{i,t}) \right] \\
\text{subject to} \quad & I(\pi_i, s_{i,t}, q_i) < \Bar{I}
\end{aligned}
\end{equation}
where agents maximise $U$ while adhering to a constraint $\Bar{I}$ on their processing costs $I$. The processing cost restricts how far the learnt policy $\pi_i$ can diverge from the prior belief $q_i$. We can equivalently reformulate \cref{eqOriginal} as the maximisation of a modified reward:
\begin{equation}\label{eqRewardMod}
    \pi_i^\lambda(a|s_i) = \max_{\pi_i}\mathbb{E}_{\pi_i}\left[\sum_{t=0}^\infty\gamma^t \left( U(a_t | s_{i,t}) - \lambda I(\pi_i, s_{i,t}, q_i) \right)\right]
\end{equation}
where $\lambda$ controls the strength of regularisation, modulating the \textit{boundedness} (or skill) of the agent. As $\lambda \to \infty$, the agent is entirely driven by their prior beliefs (performing no strategic reasoning), whereas as $\lambda \to 0$, the agent is unbounded and approximates rational behaviour. \cref{eqRewardMod} makes the formulation general and compatible with existing RL algorithms without requiring in-depth modification to the loss or optimisation process.

\subsubsection{Processing Costs}

Quantifying information processing costs in a generalised manner is desirable, as this enables compatibility with existing optimisation algorithms. Following recent achievements in constraining agent decision-making using information-theoretic costs \cite{Evans2023}, we adopt a similar approach. This information-theoretic treatment abstracts the underlying causes of such constraints, allowing a focus on learning behaviour without necessitating an in-depth understanding of the specific psychological factors at play. From an optimisation standpoint, this is advantageous, as the process remains independent of the particular details of how decisions are formulated \cite{sims2003implications}.

One of the most common information-theoretic constraints is an entropy constraint, e.g.  $H(\pi_i) = - \sum p(\pi_i) * \log(p(\pi_i))$, restricting deviations from uniform behaviour. For example, (the logit form of) QRE can be seen as maximisation under an entropy constraint. However, much research has shown the usefulness of incorporating arbitrary prior beliefs (not just uniform) \cite{evans2021maximum}, motivating extensions that measure the divergence from an arbitrary prior distribution based on the Kullback-Leibler (KL) divergence $\DKL$ \cite{ortega2013thermodynamics}.

KL regularisation has shown success in relevant domains \cite{sokota2022unified, jacob2022modeling} for formally capturing these costs. For example, \cite{jacob2022modeling} demonstrates the usefulness of a penalty for minimising $\DKL$ from expert policies in tree search and \cite{sokota2022unified} analyses $\DKL$ in a RL context for improving convergence in two-player games. However, neither of these considers the domain we propose here for better capturing human-like play in complex multi-agent social systems.

Specifically, we propose using the following information processing costs
\begin{equation}\label{eqKL}
    I(\pi, s, q) = \DKL(\pi \parallel q) = \sum_{a \in A} \pi(a|s) \log \frac{\pi(a|s)}{q(a|s)}
\end{equation}
to constrain $\pi_i$ from diverging too far from agents' prior beliefs $q$ at each state, limiting their strategic abilities. \cref{eqKL} can also be seen as equivalent to enforcing an $H$ constraint when assuming the prior beliefs are uniform (making connections with QRE, as discussed in \cref{secDiscussion}). Additionally, with this representation, the contribution of a specific action $a$ to the divergence can be identified, e.g.,

\begin{equation}
    I_a(\pi, s, q) = \pi(a|s) \log \frac{\pi(a|s)}{q(a|s)}
\end{equation}
meaning the adjustment to the reward function in \cref{eqRewardMod} can be directly linked to $a$ rather than $\pi_i$, i.e.,

\begin{equation}
    \pi_i^\lambda(a|s) = \max_{\pi_i}\mathbb{E}_{\pi_i}\left[\sum_{t=0}^\infty\gamma^t \left( U(a_t | s_{i,t}) - \lambda I_{a_t}(\pi_i, s_{i,t}, q_i) \right)\right]
\end{equation}
which is advantageous for optimisation purposes. We use this formulation throughout. As we sample more actions from this policy, we would approximate $\DKL$ as $I(\pi, s, q) = \sum_{a \in A}  I_a(\pi, s, q)$. 

In RL, information-theoretic regularisation is often employed to enhance the convergence or robustness of algorithms. For instance, Proximal Policy Optimization (PPO) utilises a $\DKL$ penalty term to prevent excessively large changes in the policy during training steps and improve the convergence. Similarly, the Soft Actor-Critic (SAC) algorithm employs $\DKL$ in its policy improvement step, limiting divergence from the previous Q-function \cite{haarnoja2018soft}. Moreover, Maximum Entropy RL introduces an entropy term to enhance exploration, convergence, and robustness \cite{eysenbach2019if, eysenbach2022maximum}. In contrast, our approach restricts divergence from an arbitrary prior belief $q$ to reflect the constraints in information processing present during human decision-making rather than being aimed at improving the algorithm's convergence. These prior beliefs $q$ (also called "magnets" \cite{sokota2022unified} or "anchors" \cite{jacob2022modeling}) may change throughout training and inference (e.g. with updated information) and can take many forms, for example, demonstrating bias towards certain actions, encoding heuristics, averaging over past decisions, or preferring historically well-performing actions.

\subsubsection{Heterogeneous Behaviours}

Effectively learning behaviours that capture the diversity of the population's decision-making is crucial for integration into agent-based simulations.

Two initial approaches could be employed for capturing heterogeneous skills of agents: Firstly, learning optimal (homogenous) behaviours $\pi^*$ as in Eq. (1) and applying heterogeneous bounds at inference, e.g., $\pi_i = \pi^* + \eta_i$, where $\eta_i$ is a noise term. Secondly, individual learning with heterogeneous $\lambda$'s, i.e., $\pi_i = \max_{\pi}\mathbb{E}_{\pi}\left[\sum_{t=0}^\infty\gamma^t \left( U(a_t | s_{i,t}) - \lambda_i I(\pi, s_i, q_i) \right)\right]$. In the following section, we describe why these approaches are insufficient before proposing an alternative that overcomes these limitations.

\paragraph{Post-hoc bounds at inference} One potential approach involves optimising $U$ without including information processing costs during training (eliminating boundedness) to learn $\pi^*$. Subsequently, this boundedness parameter is applied only to the learned policies during inference $\pi_i = \pi^* + \eta_i$, where $\eta_i$ is the noise term. For example, \cite{campanaro2023roll} applies dropout during simulation, and noisy introspection applies noise into the decisions relaxing the equilibrium requirement \cite{goeree2004model}. Such methods introduce a range of skill levels in action execution, e.g. through introducing noise $\eta_i$ into the action selection process. However, if heterogeneous bounds were implemented in this manner, agents would not learn how to adapt to the behaviour of other bounded agents, as the best response to the optimal policy is not necessarily the best response to a noisy policy $BR(\pi^*) \not\equiv BR(\pi_i)$. 
To illustrate this point, consider a simple rock-paper-scissors (RPS) environment. In RPS, the perfectly rational equilibrium policy is $\pi^*= \{p(\text{R}), p(\text{P}), p(\text{S})\} = \{\frac{1}{3}, \frac{1}{3}, \frac{1}{3}\}$. However, if agent 2s policy is instead fixed as $\pi_2=\{\frac{0}{3}, \frac{0}{3}, \frac{3}{3}\}$ (e.g. they are boundedly rational and biased towards playing $S$), the rational best response for agent $1$ is $\pi_1 = BR(\pi_2) = \{\frac{3}{3}, \frac{0}{3}, \frac{0}{3}\}$. Had agent $1$ not observed $a_2 \sim \pi_2$ during training, they would not have learned to exploit $\pi_2$. While a simple example, this consideration becomes pivotal in demonstrating the emergence of auto-curricula \cite{Baker2020Emergent}. To capture the interplay among heterogeneously skilled agents, the notion of boundedness must be present throughout the learning process rather than only during inference. 

\paragraph{Individual Learning} An alternative approach involves assigning heterogeneous processing penalties $\lambda_i$ to each agent $\pi_i=\pi^{\lambda_i}$ in \cref{eqRewardMod}, e.g. with heterogeneous logit responders \cite{golman2011quantal}. In this scenario, any standard MARL algorithm could be employed, wherein all agents, each governed by their unique constraint, strive to optimise their rewards, adjusting their behaviours in response to the observed outcomes. However, this method would prove inefficient due to the necessity of learning $N$ individual policies and calibrating $N$ different individualised processing costs $\lambda_i$ (one for each agent). This inefficiency becomes a crucial concern as ABMs often have a large $N$. Additionally, the learnt policies would not generalise across different $\lambda_i$, requiring new training each time a new $\lambda_i$ is introduced.

It becomes clear we need a scalable alternative that can deal with the heterogeneous boundedly rational behaviour of agents.

\subsection{Shared Policy Learning}

Rather than learning individual policies with heterogeneous $\lambda_i$, all aiming to solve \cref{eqRewardMod}, we wish to learn a generalised policy $\pi(\dots|s_i, i, \lambda_i, q_i)$. This representation treats agents' prior beliefs and processing resources as part of the observation space (and for simplicity of notation, we will use $\pi_i(\dots|s_i) = \pi(\dots|s_i, i, \lambda_i, q_i)$), enabling generalised policy learning based on these state observations. This formulation provides a way of efficiently representing a diverse population of agents $i \in N$ with varying bounds in strategic reasoning abilities $\lambda_i$ through a single parameterised policy with an augmented observation space.

However, calibrating $\lambda_i$ remains an important issue. While calibrating $\lambda_i$ to each agent $i$ may seem ideal, this is computationally impractical with larger $N$, and additionally, could lead to overfitting to specific behavioural parameters due to the large number of required parameters ($N$). Furthermore, in practice, as $\lambda_i$ are unobserved, assigning these values exactly is difficult.

We adopt an alternative approach to address this challenge by assigning individual strategic processing resources as samples from a probability distribution $\lambda_i \sim \mathcal{D}$, addressing the uncertainty of the agents' exact $\lambda_i$ values and keeping the number of free parameters low. Any $\mathcal{D}$ could be utilised (and the proposed approach is agnostic to the particular distribution used), but here, we employ the Gaussian distribution $\mathcal{D}=\mathcal{N}(\mu,\sigma)$, where $\mu$ controls the mean processing costs, and $\sigma$ the heterogeneity. This way, we only need to calibrate the parameters $\mu$ and $\sigma$ (rather than $N$ separate parameters), which is typically $<<N$. For instance, when dealing with $N=100$ agents, we are calibrating just $2$ parameters instead of $100$, helping to avoid overspecification. During learning, this sampling approach allows for interpolation across a range of $\lambda_i$, reducing the computational complexity and enforcing a "smoother" policy. This smoothness arises from observing many different behaviours during training, resulting in a more robust policy forced to interpolate across $\lambda_i$ values, reducing the potential of overfitting.

\begin{figure}[!htb]
    \centering
    \includegraphics[width=.97\columnwidth]{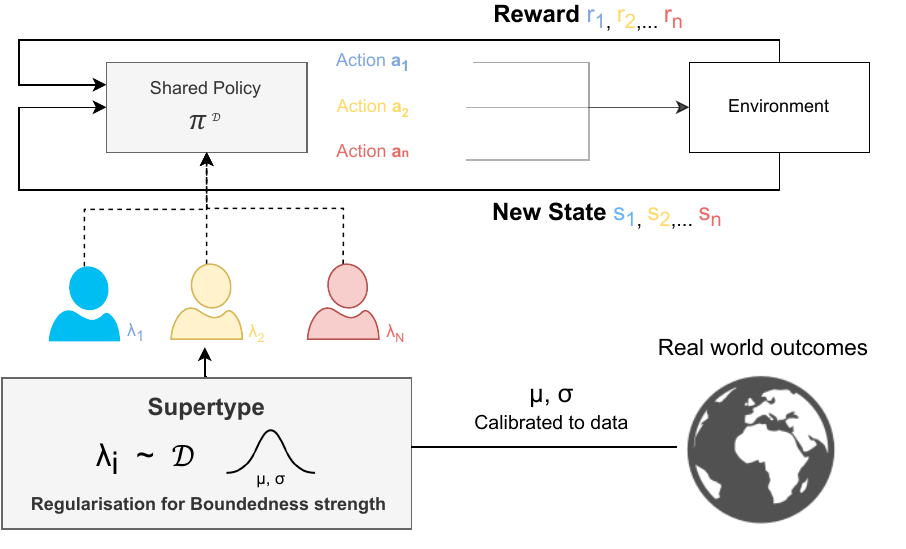}
    \caption{Proposed Approach: Shared policy learning with heterogeneous bounds through agent supertypes.}
    \label{figDiagram}
\end{figure}

To learn a generalised policy $\pi^\mathcal{D}$ for $\lambda_i \sim \mathcal{D}$, we use agent supertypes \cite{vadori2020calibration}, enabling efficient scaling through shared policy learning, while still capturing a range of behaviours. Agent supertypes have exhibited promise, particularly in applications such as calibrating rational behaviour in over-the-counter markets \cite{vadori2022towards}. However, the potential for incorporating heterogeneous strategic reasoning skills to better approximate human decision-making has yet to be explored. Here, we propose a novel approach to extend the use of supertypes to capture diverse bounded rational behaviours based on the regularised policies introduced in \cref{eqRewardMod}. 

Under the proposed approach, a regularised policy for the supertype is established $\pi^\mathcal{D}$, which is exposed to different regularisation strengths $\lambda_i \sim \mathcal{D}$ throughout training. $\pi^\mathcal{D}$ learns to extrapolate over the regularisation strengths, reducing the number of policies to train while still enabling heterogeneous behaviour. Through this process, agents learn to adapt their behaviour in response to the varying processing resources across the agent population, accounting for potential auto curricula. The inputs to the supertype $(\mu, \sigma)$ are calibrated such that that the simulation outcome closely matches the real world dynamics (from the calibration data). An overview of this high-level process is depicted in \cref{figDiagram}. 

The shared policy $\pi^\mathcal{D}$ takes an agent's id $i$, processing resources $\lambda_i$, and prior beliefs $q_i$ as inputs (as components of $s_i$). Including $i$ facilitates the learning of (potentially) competitive behaviour between agents of the same supertype. By adjusting the input $\lambda_i$ values, $\pi^\mathcal{D}$ can effectively demonstrate a spectrum of skill levels, while only needing to learn a single (generic) policy. Allowing for arbitrary $q_i$ accounts for the effect of various prior beliefs. 

The underlying assumption of this supertype approach is that all agents in the supertype have the same $U$ function; however, they possess varying levels of skill in maximising $U$. Given the inherent uncertainty surrounding the precise nature of agent decision-making, a compelling case is made for capturing a spectrum of regularised behaviours. The regularisation offers dual advantages: firstly, it enables deviations from perfect rationality in agent behaviour; secondly, it effectively encompasses uncertainties from both the modeller's perspective and the agents being modelled (i.e., uncertainty in the model's formulation and the agents' decision processes) \cite{evans2021maximum}. This motivation aligns with the use of bounded rationality in situations characterised by fundamental uncertainty of the agent \cite{gigerenzer2020bounded} and also helps to address concerns regarding modeller judgement (e.g. model misspecification) by permitting a range of information-constrained behaviour \cite{scharfenaker2020implications}.

\section{Empirical Results: $n-$agent Settings}\label{secResults}

To verify that the proposed approach can capture a range of interesting behaviour not predicted by the analytically derived equilibrium or standard state-of-the-art MARL approaches, we compare the predictions from the proposed model against these approaches on a range of canonical $n$-agent economic environments involving human participants.

\begin{table}[!htb]
\centering
\caption{$5x2$-fold validation results for each environment. Each cell displays the root mean squared errors as mean $\pm$ standard deviation, along with (rankings) for between-environment comparison \cite{demvsar2006statistical}. The last row presents the average rank\protect\footnotemark. Lower rankings indicate better performance. }\label{tblResults}
\adjustbox{width=\columnwidth}{%
\begin{tabular}{@{}llll@{}}
\toprule
                      & Rational & MARL & \textbf{Proposed} \\ \midrule
Supply Chain          &  0.33 $\pm$ 0.004 (2.5)                 &             0.33 $\pm$ 0.004 (2.5)                                                                               &   0.02 $\pm$ 0.005 (\textbf{1})  \\[3pt]
Cournot          &                          &                                                                                          &                                                                                                \\

- Duopoly         &   0.16 $\pm$ 0.001 (3)                   &                                                                                
0.13 $\pm$ 0.001 (2)
         &     0.04 $\pm$ 0.001  (\textbf{1})                                                                                        \\
- Triopoly          &    
0.16  $\pm$ 0.002 (3)
&  0.15 $\pm$ 0.002 (2) &  0.03 $\pm$ 0.001 (\textbf{1})  \\[3pt]
Cobweb         &   0.02 $\pm$ < 0.001  (2)   &  0.03 $\pm$ < 0.001 (3)   &    0.01 $\pm$ < 0.001 (\textbf{1})                                                                                           \\ \midrule
\textbf{Rank} &           2.5               &                                                                            2.5              &   \textbf{1}                                                                                         \\ \bottomrule
\end{tabular}
}
\end{table}
\footnotetext{The Cournot competition environments each carry $\frac{1}{2}$ weight when computing the average to address the interdependence and avoid biasing the average ranking (although in this case, the rankings would not change without such a weighting)}

\subsection{Process Overview}
We assess the performance of the proposed approach in three well-established multi-agent economic environments: supply chains, oligopolies, and cobweb markets. To validate our approach, we leverage laboratory experiments conducted in each setting, comparing the predictions with actual human behaviour. We compare the proposed approach with analytically derived solutions and a state-of-the-art MARL algorithm (PPO). In each case, we perform repeated 5x2cross-fold validation \cite{demvsar2006statistical} to estimate the generalisation ability, ensuring the models do not overfit to the calibration data. The squared $L2$-loss function (the mean squared error) is used as our performance metric \cite{d2023loss}. In the presented tables, we report the root mean squared error for interpretability. To facilitate comparisons across environments, we use rankings based on resulting errors (where the lowest error receives rank=1, and ties are split by using the average rank had there been no ties) \cite{demvsar2006statistical}. To calibrate our model, we perform a grid search over $\mu, \sigma$ values, choosing $\mu, \sigma$ with the lowest training error for use on the unseen test set. The optimisation process never sees the test data. Additional experiment details are given in \cref{appendixExperiment}.

\subsubsection{Calibration}

\begin{figure}
\begin{subfigure}{.4\columnwidth}
\includegraphics[width=\textwidth]{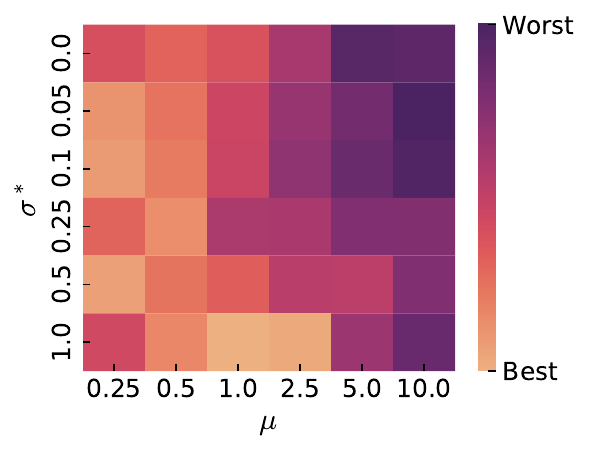}
    \caption{$\mu$, $\sigma*$}
\end{subfigure}
\begin{subfigure}{.5\columnwidth}
\includegraphics[width=\textwidth]{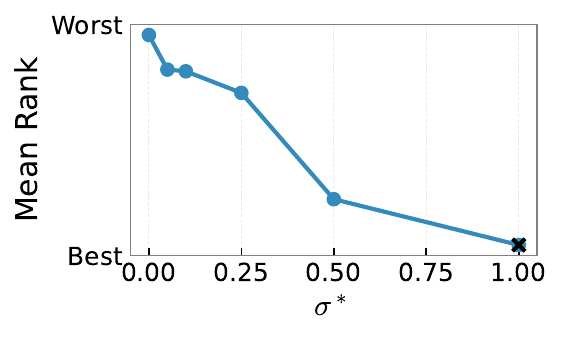}
    \caption{$\sigma^*$ averaged across values of $\mu$}\label{figTriopolyCalibrationResultsSigma}
\end{subfigure}
    \caption{Triopoly calibration results for values of the boundedness parameter $\mu$ and heterogeneity parameter $\sigma^*$}
    \label{figTriopolyCalibrationResults}
\end{figure}

The calibration results for one environment (triopolies), displaying the values of the heterogeneity ($\sigma^*$) and boundedness ($\mu$) parameters, are shown in \cref{figTriopolyCalibrationResults}. Similar plots are available for all environments in \cref{appendixCalibration}. The proposed approach offers the flexibility to incorporate perfect rationality or homogeneity by setting $\mu=0$ (removing bounds) or $\sigma^*=0$ (removing heterogeneity). However, it's noteworthy that the optimal values never align with $\mu=0$ or $\sigma^*=0$, highlighting the usefulness of both heterogeneity and processing costs across all three environments. Furthermore, since homogeneity and unbounded reasoning can be considered special cases of our proposed approach, this eliminates the need to determine such assumptions \textit{a priori} (as often required in many existing methods). Instead, our approach enables the calibration of these properties based on the specific environments of interest.

\subsection{Results}
The out-of-sample performance of each algorithm is compared in \cref{tblResults}. The proposed approach achieves the highest accuracy across all three environments, resulting in the best overall rank. The state-of-the-art (standard) MARL approach and the analytically derived rational case generally perform equivalently, indicating that the MARL algorithm approximated the true rational equilibrium well. However, both the alternatives performed poorly in capturing the experimental data, demonstrating that rationality and homogeneity are too strict of an assumption even in these relatively simple multi-agent settings. These results motivate the relaxation of perfect rationality and the introduction of skill heterogeneity when using MARL to model complex systems.

To better understand the results and the reason for the improved capabilities of the proposed approach, we analyse each environment in more detail. For each environment, we begin with a brief description, before presenting the results.

\subsubsection{Supply Chains}

\paragraph{Description} The supply chain environment is a capacity allocation problem with a single good with cost $c$ and price $p$. There is one supplier with a limited capacity $K$, and $I$ retailers. Each retailer $i$ makes a request $0 < x_i \leq X, x_i \in \mathbb{Z}$, and the supplier responds by offering $y_i$. Retailers are allocated goods proportionate to their request $y_i \propto x_i$:

\begin{equation}
    y_i = K \times \frac{x_i}{\sum_{j \in I} x_j}
\end{equation}\label{eqProportion}inducing the potential for (rationally) inflated order sizes to ensure the required quantities are met.
Each retailer receives a fixed demand $D > \frac{K}{I}$, i.e., resources are limited. The reward is given by

\begin{equation}
    U(x_i) = D \times (p-c) - \omega \times \max(y_i-D, 0) - s \times \max(D-y_i, 0)
\end{equation}
where $\omega$ is the wastage cost, and $s$ is the shortage cost ($s=p-c$). The rational (Nash) solution to this task, irrespective of $\omega, s$ (when in the limited capacity case of $K<I*D$), is for retailers to submit their maximum request $X$, each retailer then receiving $y_i=\frac{K}{I}$ units due to the proportional allocation. Any lower of a request would result in the retailer receiving $y_i < \frac{K}{I} < D$.

We utilise the experimental results of \cite{doi:10.1287/mnsc.1120.1531}, with $I=2$, $\omega=2$, and $s=5$. There were $30$ subjects, composed of university students, randomly paired in $30$ repeated decision rounds to make a game with two retailers in each round. The capacity is $K=90$, with each retailer receiving demand $D=50$ and able to make a maximum request of $X=100$. As $2D>K$, we are faced with limited capacity. 

\begin{figure}[!htb]
    \centering
    \includegraphics[width=.75\columnwidth]{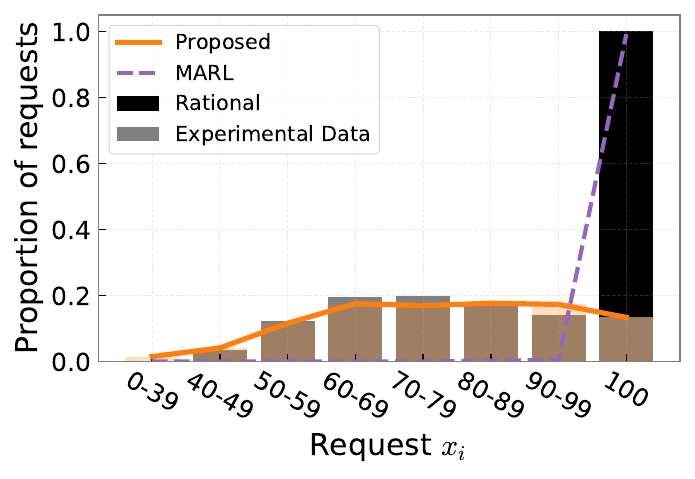}
    \caption{Supply Chain. Experimental data from \cite{doi:10.1287/mnsc.1120.1531} are shown as grey bars. The proposed approach is shown with the orange line (for one calibration fold). The standard MARL approach is shown as the dashed purple line, and the NE is denoted by the black bar.}
    \label{figSupplyChain}
\end{figure}

\paragraph{Results} The experimental results are displayed in \cref{figSupplyChain}, showing substantial deviations from purely rational (Nash) play. The NE is to request the maximum $x_i = X = 100$. The standard MARL approach learns the NE here; however, this is a poor predictor of what happens experimentally (\cref{figSupplyChain}). Experimentally, the most commonly occurring requests are in the $60-80$ range, far lower than the NE. The proposed approach is a very good fit for the experimentally observed behaviour, capturing the overall trend, demonstrating that subjects have varying strategic bounds, giving rise to a range of outcomes not predicted by a rational representative agent, and helping to motivate the bounded rationality assumptions.

\subsubsection{Cournot Oligopoly}

\paragraph{Description} The Cournot competition is an environment modelling oligopolies in a market. In a Cournot market, $K$ firms must simultaneously choose what quantities $q_i \in \mathbb{Z}$ of a homogenous good to produce. The reward for firm $i$ depends on the market price $p$ of the good and the individual  $q_i$, i.e.:

\begin{equation}
    U_i = p \times q_i
\end{equation}
where $p$ is determined by the total production of all goods, i.e.,

\begin{equation}
    p = A - B \times \sum_{k=1}^K q_k
\end{equation}

We use the experimental data from \cite{ho2021bayesian, fouraker1963bargaining}, for duopolies and triopolies (experiments 7,8,9,10 from \cite{fouraker1963bargaining}). Following \cite{ho2021bayesian}, we group experiments $7,10$ (duopoly) together and experiments $8,9$ (triopoly) together. In these experiments, $A=2.4$, $B=0.04$, and $8 \leq q_i \leq 32$. There were $64$ participants for the duopoly experiments, and $66$ for the triopoly, composed of university students.

\begin{figure}[!htb]
\begin{subfigure}{.45\columnwidth}
    \centering
    \includegraphics[width=\textwidth]{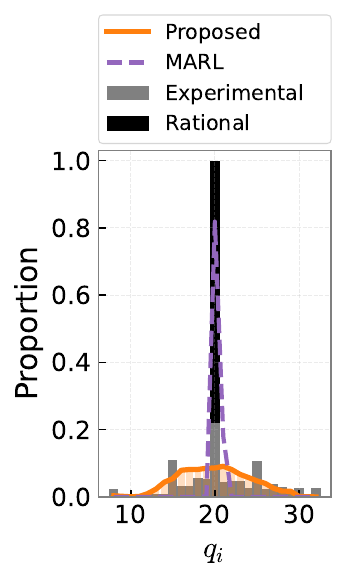}
    \caption{Duopoly}
    \label{figDuopoly}
\end{subfigure}
\begin{subfigure}{.45\columnwidth}
    \includegraphics[width=\textwidth]{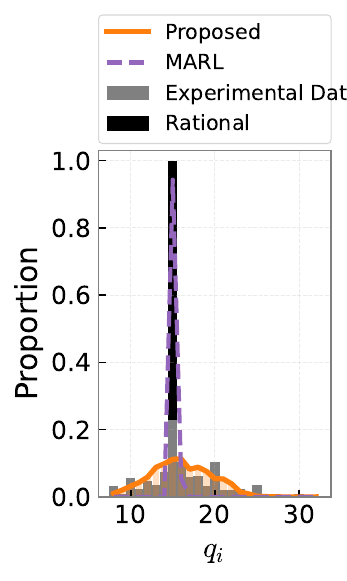}
    \caption{Triopoly}
    \label{figTriopoly}
\end{subfigure}
\begin{subfigure}{\columnwidth}
\centering
\includegraphics[width=.9\textwidth]{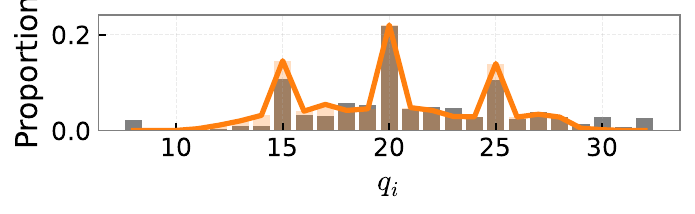}
\caption{Duopoly with \textit{a priori} preference towards prominent numbers ($10, 15, 20, 25, 30$), reflecting a cognitive bias.}
\label{figPriors}
\end{subfigure}

\caption{Cournot competitions. Experimental data from \cite{fouraker1963bargaining} is shown as grey bars. The proposed approach is shown with the orange line (for one calibration fold). The standard MARL approach is shown as the dashed purple line, and the NE is denoted by the black bar.
}\label{figOligopoly}
\end{figure}

\paragraph{Results} The results for duopolies (triopolies) are presented in \cref{figDuopoly} (\cref{figTriopoly}). With the experimental data, we see a significant deviation from both the rational behaviour and the standard MARL predictions. The unique NE for duopolies and triopolies is $20$ and $15$ respectively. While these actions are the most common in each case, these occurrences comprise $\leq20\%$ of the total decisions, and the remaining $\approx 80\%$ are sub-optimal decisions (under the assumption of mutual rationality). Again, the proposed model is a good fit for the experimental data in both duopolies and triopolies, capturing this significant deviation from the optimal choice while still capturing the maximal peak from the experimental data. 

Under the processing cost constraint, agents choose actions proportionate to the expected reward and the level of regularisation in their decision function (their skill level). This means there are specific over representative peaks in the experimental data, for example, at $15$ and $25$ in the duopoly case and $20$ in the triopoly case, which the model with uniform priors can not capture. These peaks can not be explained from expected reward alone, as there is no particular reason that $15$ would have such high preference. Instead, these demonstrate an \textit{a priori} preference of the agents towards particular prominent numbers ($0, 5, 10, \dots$), a known cognitive bias \cite{converse2018role, chen2018round}. Owing to the model's flexibility in allowing for arbitrary prior beliefs, this can be modelled with $q_i$ with higher weightings on these prominent numbers. An example of the resulting decisions when using such priors is shown in \cref{figPriors}, providing a significantly improved fit, capturing all of the experimental peaks. We do not use such a model when comparing results in \cref{tblResults}, as this modification was made post-hoc (after seeing the experimental data), but it shows the usefulness of incorporating prior beliefs when known, demonstrating an additional strength of the model.

\subsubsection{Cobweb Market}

\paragraph{Description} In a cobweb market \cite{hommes2007learning}, there are $K$ producers who must estimate the price $\hat{p}_{i,t}$ of a good at the next timestep $t$. The reward for a producer $i$ is based on the accuracy of their prediction compared to the market price $p_t$:

\begin{equation}
    U_{i,t} = \max(0, 1300-260(p_t - \hat{p}_{i,t})^2)
\end{equation}
which is lower bound by $0$, e.g., the producers cannot receive negative utilities. Producers have no contact with others, but at the end of each round, producers observe the realised market price $p_t$.

The market price depends on the demand $D$ and supply curves $S$. $D$ is linear with price and is subject to small normally distributed demand shocks $\eta_t$, and $S$ non-linearly increases with the producer's expected price, i.e., 
\vspace{-1mm}
\begin{equation}
\begin{aligned}
    D(p_t) &= a - bp_t + \eta_t \\
    S(\hat{p}_{i,t}) &= \tanh (\psi (\hat{p}_{i,t} - K)) + 1
\end{aligned}
\end{equation}
where $\psi$ controls the non-linearity and stability of the market. The realised market price is given by

\begin{equation}
    p_t = \frac{a - \sum_{k \in K} S(\hat{p}_{k,t})}{b} + \epsilon_t
\end{equation}
where $\epsilon_t \propto \eta_t$. Under rational expectations, producers all predict the price to be the intersection of $S$ and $D$, $p*$, e.g., 
$\bar{p}_{i,t} = p* + \epsilon_t, \forall i \in K$, meaning the rational predictions will, on average, fall in line with the equilibrium price with fluctuations $\propto \epsilon_t$.

We utilise the experimental data of \cite{hommes2007learning}, with $a=13.8$, $b=1.5$, $\epsilon_t \sim \mathcal{N}(0,0.5)$ and $\psi=2$. There were 36 participants, generally undergraduate economics, psychology, and science students.

\paragraph{Results}

\begin{figure}[!htb]
    \centering
    \includegraphics[width=.9\columnwidth]{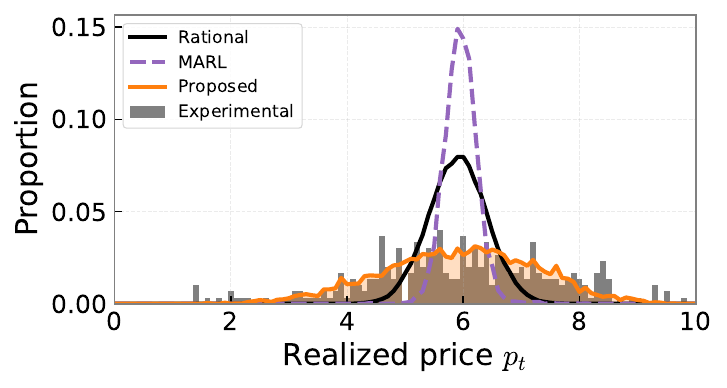}
    \caption{Distribution of $p_t$ in cobweb markets. Experimental data from \cite{hommes2007learning} is shown as grey bars. The proposed approach is shown with the orange line (for one calibration fold). The standard MARL approach is shown as the dashed purple line, and the black line denotes the rational expectations solution.}
    \label{figCobweb}
\end{figure}

The cobweb market results are visualised in \cref{figCobweb}, displaying the distribution of realised prices. Both the standard MARL approach and the rational expectations are poor predictors of the observed phenomena from the experimental data. While the mean of the experimental data often aligns with the rational and MARL case, the distribution spread is far broader, indicating persistent excess volatility, with much larger standard deviations than those expected under the rational expectations hypothesis or MARL approach. This is a noteworthy stylised fact of markets incompatible with the rationality assumption of all agents \cite{hommes2007learning, hommes_lux_2013}. As excess volatility is known to occur in many markets \cite{vyetrenko2020get}, understanding the causes and being able to model this volatility is an important use of ABM. The proposed approach offers a much better fit, capturing the mean of the data and the overall distribution of price fluctuations, reproducing the observed excess price volatility (\cref{figCobweb}), providing an explanation of the endogenous formation of excess volatility based on bounded rationality.

\subsection{Key Takeaways}
The proposed approach demonstrated strong out-of-sample performance across these three economic and financial environments, outperforming the comparisons and validating the model in controlled environments. Specifically, we showcased the value of:
\begin{itemize}
\item \textbf{Boundedness}: Incorporating bounded rationality resulted in substantially improved predictive accuracy (\cref{tblResults}).
\item \textbf{Heterogeneity}: Allowing for heterogeneous processing costs improved upon assuming mutual consistency (\cref{figTriopolyCalibrationResultsSigma})
\item \textbf{Non uniform priors}: Arbitrary prior beliefs explained phenomena incompatible with deviations from expected utility alone (\cref{figPriors})
\end{itemize}
Additionally, while the proposed approach relaxes these three assumptions, if desired, these can still be recovered as special (limit) cases as discussed in \cref{secDiscussion}. The benefit of the proposed approach is that these assumptions do not need to be established \textit{a priori}, rather they are calibrated to the environment of interest.

\section{Discussion and Relation to equilibrium solutions}\label{secDiscussion}
Flexibility is one of the model's strengths. However, this flexibility comes at the expense of exact analytical tractability, and generally, we are limited by the theoretical guarantees of the underlying RL algorithm (here, PPO). Despite this, in this section we show the relation to the decision functions of other equilibrium solution concepts and provide discussions on the equilibrium approximations.

\paragraph{Quantal Response Equilibrium}
With homogeneous processing costs $\lambda_i=\lambda$ and uniform prior beliefs $q_i(a) = q$, the approach can be seen as approximating QRE (as QRE converges to NE with $\lambda \to 0$ \cite{goeree2020stochastic}, approximation of NE too). Using a similar formulation to \cref{secFormulation}, with QRE, each agent chooses $a$ to maximise $U$, subject to an entropy $H$ constraint:

\begin{equation}
\begin{aligned}
& \max {\pi_i}(a) U(a|\pi_{-i}) \\
\text{subject to} \quad & H(\pi_i) \geq H_{\text{min}}
\end{aligned}
\end{equation}
where $\pi_{-i}$ gives the action profile of the other agents. To derive the quantal response decision function $\text{QR}_i$, we use the method of Lagrange multipliers and the principle of maximum entropy to convert this into an unconstrained optimisation problem. Given the usual constraints on the probability function (that $\text{QR}_i(a) \geq 0, \forall a$ and $\sum_{a \in A}\text{QR}_i(a) = 1$), we get the following Lagrangian \cite{evans2021maximum}:

\begin{equation}\label{eqLagrange}
    \mathcal{L} = - \sum_{a \in A} \text{QR}_i(a)U(a|\text{QR}_{-i}) - \zeta  \left( \sum_{a \in A} \text{QR}_i(a) - 1 \right) + \lambda \Biggl( H(\text{QR}_i) - H_{\text{min}} \Biggr)
\end{equation}
where taking the first order conditions and solving for $\text{QR}_i$ yields

\begin{equation}\label{eqQRE}
    \text{QR}_i(a) = \frac{e^{U(a|\text{QR}_{-i}) / \lambda }}{\sum_{a' \in A} e^{U(a'|\text{QR}_{-i}) / \lambda }}
\end{equation}

To demonstrate that the decision function implied by the $\DKL$ constraint in \cref{eqKL} (with uniform priors and homogenous $\lambda$) reduces to the same functional form as \cref{eqQRE},  we get:

\begin{equation}
    I(\pi, s, q) = \sum_{a \in A} \pi(a|s) \log \frac{\pi(a|s)}{q(a|s)} = \sum_{a \in A} \pi(a|s) (\log (\pi(a|s)) - C) \\
\end{equation}
plugging into $\mathcal{L}$
\vspace{-1mm}
\begin{equation}\label{eqLagProposed}
\begin{aligned}
\mathcal{L} = - \sum_{a \in A} \pi_i(a|s_i)U(a|\pi_{-i}) - \zeta  \left(\sum_{a \in A} \pi_i(a|s_i) - 1\right) + \\ \lambda \Biggl(\sum_{a \in A} \pi(a|s) (\log (\pi(a|s)) - C) - \bar{I}\Biggr)
\end{aligned}
\end{equation}
and the decision function reduces to:
\begin{equation}
    \pi_i(a|s_i) = \frac{C e^{U(a|s_i) / \lambda }}{\sum_{a' \in A} C e^{U(a'|s_i) / \lambda }} = \frac{e^{U(a|s_i) / \lambda }}{\sum_{a' \in A} e^{U(a'|s_i) / \lambda }}
\end{equation} 
confirming equivalent functional forms to \cref{eqQRE} under uniformity and homogeneity. The key difference is $\text{QR}_i$ depends directly on the policies of other agents $\text{QR}_{-i}$, whereas $\pi_i$ captures this indirectly via the state $s_i$.

The QRE then corresponds to a fixed point of these QR functions \cite{goeree2020stochastic}, assuming that $\lambda$ is homogeneous and common knowledge among the agents. In contrast, under the proposed approach, rather than explicitly attempting to find the fixed point solution, gradient descent and simulation are used to find $\pi_i$  that maximises $U$, with a neural network $f$ (with inputs $s_i$ including $q_i, \lambda_i$), and no common knowledge of $\lambda_j, q_j, j \neq i$. The outputs of $f$ are $|A|$ logits (one for each $a \in A$), which are passed through a softmax function, giving learnt policies of the form:
\vspace{-1mm}
\begin{equation}
    \hat{\pi}_i(a) = \frac{e^{f_i(a|s_i)}}{\sum_{a' \in A} e^{f_i(a|s_i)}}
\end{equation}
where each agent is continually attempting to learn $f_i$ that maximises their expected reward from using $\hat{\pi_i}$ (here using PPO with Generalized Advantage Estimation \cite{schulman2015high}). Of course, except in very specific settings \cite{zhang2019policy}, we do not have general convergence guarantees, so we say the proposed approach \textit{approximates} these equilibria. 

The benefit of the proposed approach is the flexibility of $f$ in allowing for various behaviours from heterogenous agents (e.g., varying $\lambda_i$ and $q_i$) and computability when deriving the equilibria would otherwise be intractable, such as when $\lambda_i$ and $q_i$ are not common knowledge. When allowing heterogeneous $\lambda_i$ and $q_i$, we approximate a Subjective Heterogeneous Quantal Response Equilibrium \cite{rogers2009heterogeneous}, a type of Bayesian equilibrium \cite{geanakoplos1994common}, where agents may have different (potentially incorrect) subjective beliefs about the type distributions of the other agents (in this case, the values of  $\lambda_i$ and $q_i$ in the population).

\paragraph{Relation to Rational Inattention}

As mentioned in \cref{secBackground}, the key relevant work in this area is \cite{mu2022modeling}. While we share a similar goal, our work differs in some important ways. \cite{mu2022modeling} requires estimating the mutual information (MI) for processing costs using a separate estimation engine. MI is defined over the joint probabilities as:
\vspace{-1mm}
\begin{equation}
    MI = - \sum_{a \in A} p(a,s_i) \log \frac{p(a,s_i)}{p(s_i)p(a)}
\end{equation}
which has a dependence on the unconditional $p(a)$ which must be solved with approximation techniques \cite{evans2021maximum}. As the divergence we utilise in \cref{eqKL} does not have this same dependence, such estimation is not required, providing an alternative formulation allowing for arbitrary prior beliefs $q_i$, useful for representing cognitive biases (as demonstrated in \cref{figPriors}) or encoding behavioural heuristics. Furthermore, as discussed, we allow for a range of heterogeneous agent skills learnt through regularised policies, and propose an approach for efficiently calibrating these policies with agent supertypes and shared policy learning, both yet to be considered. 

\section{Conclusions}\label{secConclusions}
Agent-based models have much promise for explaining complex phenomena in a broad range of disciplines. However, a key criticism is how the behavioural rules are defined. Learning realistic behavioural rules calibrated to real-world systems is essential to improve the models and promote continued uptake. In this work, we proposed an efficient MARL approach for inferring these decisions by calibrating heterogeneously skilled learning agents to real-world systems through shared policy learning and agent supertypes.

Under the proposed approach, agents possess diverse strategic processing abilities, represented through regularisation in their decision function. This regularisation is in the form of information processing costs, leading to varying levels of boundedly rational strategic behaviour, depending on the strength of regularisation. This agent skill heterogeneity is a critical aspect of many systems and is a departure from traditional equilibrium definitions. However, we demonstrate that this heterogeneity better captures many phenomena, as demonstrated under the various laboratory settings here and observed in many other real-world situations. For example, in market settings, institutional investors may have higher access to information and more extensive processing abilities than retail investors, altering the resulting market dynamics and potentially giving rise to behaviour deviating from the mutually consistent equilibrium. Relaxing this strict notion of equilibrium allows modelling a much broader range of dynamics.

\balance

The proposed approach does not impose strict assumptions on rationality, mutual consistency, or homogeneity but instead simulates the emergent outcomes through learning among the interacting agents. While these assumptions are not imposed, they can be recovered as special cases of the proposed approach, eliminating the requirement of determining which features are relevant \textit{a priori}. We evaluated the proposed approach in various economic environments, demonstrating improved out-of-sample predictive accuracy compared to existing state-of-the-art MARL methods (PPO) and analytically derived equilibrium solutions. This work provides a valuable tool for modelling complex social systems and calibrating these models to real-world dynamics, particularly when analytical approaches become intractable, setting the foundation for more advanced simulations, e.g. limit order books \cite{liu2022biased}.

\newpage
\section*{Disclaimer}
This paper was prepared for informational purposes by the Artificial Intelligence Research group of JPMorgan Chase \& Co and its affiliates (“J.P. Morgan”) and is not a product of the Research Department of J.P. Morgan.  J.P. Morgan makes no representation and warranty whatsoever and disclaims all liability, for the completeness, accuracy or reliability of the information contained herein.  This document is not intended as investment research or investment advice, or a recommendation, offer or solicitation for the purchase or sale of any security, financial instrument, financial product or service, or to be used in any way for evaluating the merits of participating in any transaction, and shall not constitute a solicitation under any jurisdiction or to any person, if such solicitation under such jurisdiction or to such person would be unlawful. © 2024 JPMorgan Chase \& Co. All rights reserved.

\balance

\bibliographystyle{ACM-Reference-Format} 
\bibliography{bib}

\clearpage
\appendix
\renewcommand\thefigure{\thesection.\arabic{figure}}    

\section{Training}\label{appendixExperiment}

\begin{figure}[!htb]
    \centering
    \includegraphics[width=.7\columnwidth]{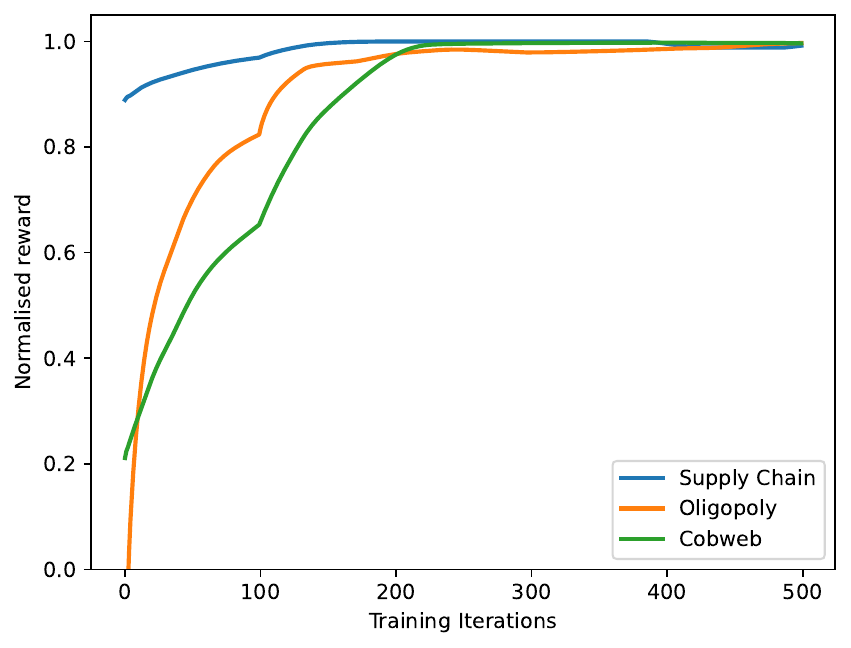}
    \caption{Training convergence}
    \label{figConvergence}
\end{figure}

Each environment is configured in Phantom \cite{ardon2023phantom}, with a RLLib backend \cite{liang2018rllib}. Agents are strategic agents, learning via PPO \cite{schulman2017proximal}, with a neural network with 2 hidden layers, of 64 nodes in each layer, and discrete ordinal discrete action spaces \cite{tang2020discretizing}. All other parameters keep their default values from RLLib. To ensure equitable comparison, the proposed approach and the standard MARL algorithm use the same hyperparameters, observation spaces, and action spaces, and the training process is executed for an identical number of iterations (500) across both approaches, ensuring ample time for convergence, as demonstrated in \cref{figConvergence}.

\subsection{Calibration}

\balance

For the proposed approach, due to the principle of insufficient reason, we assume uniform prior beliefs among the agents. Although we provide discussion with varying priors to show the flexibility of the proposed approach (e.g. \cref{figPriors}), we do not use these for comparison due to potential leaking effects from setting priors after observing data. We do not calibrate or alter the priors, but show the possibility and benefit of doing so.

We calibrate $\mu, \sigma$ from $$\mu \in \{0, 0.25, 0.5, 1, 2.5, 5, 10\}$$ $$\sigma^* \in \{0, 0.05, 0.1, 0.25, 0.5, 1\}$$ where $\sigma = \mu \times \sigma^*$. We restrict $\sigma^* \leq 1$ as we are dealing with normal distributions and do not want negative processing penalties ($\lambda_i < 0$ is clipped at $\lambda_i=0$). The calibration is the result of the lowest mean squared error on the training fold. The testing folds are never used for calibration.

\subsubsection{Calibration Results}\label{appendixCalibration}

\begin{figure}[!htb]
\begin{subfigure}{.2\textwidth}
    \centering
    \includegraphics[width=\textwidth]{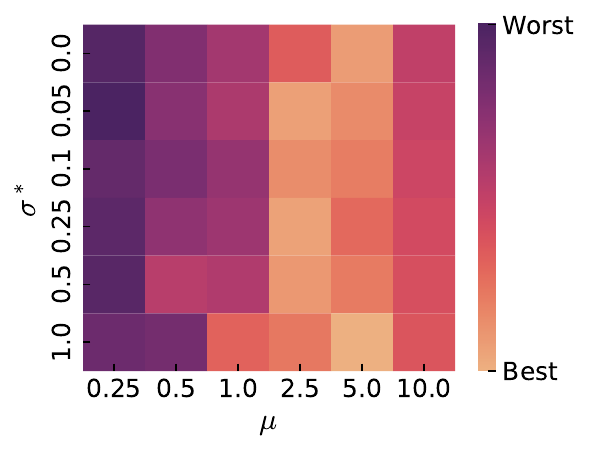}
    \caption{Supply Chain}
\end{subfigure}
\begin{subfigure}{.2\textwidth}
\includegraphics[width=\textwidth]{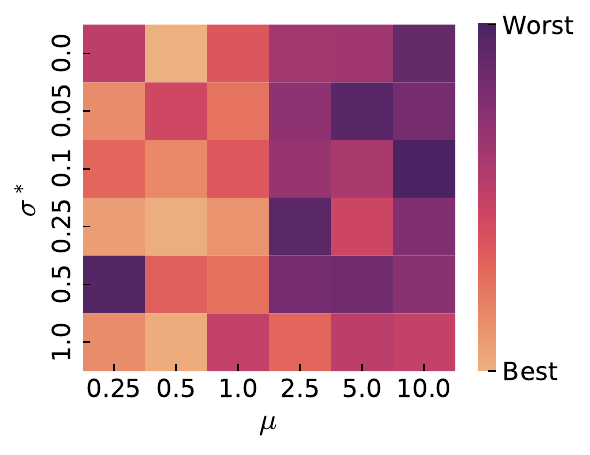}
    \caption{Duopoly}
\end{subfigure}
\begin{subfigure}{.2\textwidth}
\includegraphics[width=\textwidth]{images/calibration/triopoly_calibration.pdf}
    \caption{Triopoly}
\end{subfigure}
\begin{subfigure}{.2\textwidth}
\includegraphics[width=\textwidth]{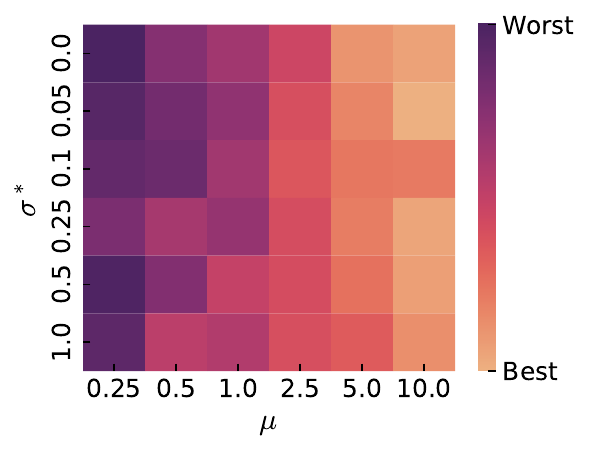}
    \caption{Cobweb}
\end{subfigure}
    \caption{Calibration Results for values of $\mu$ and $\sigma*$}
    \label{figCalibrationResults}
\end{figure}

\begin{figure}
\begin{subfigure}{.2\textwidth}
    \centering
    \includegraphics[width=\textwidth]{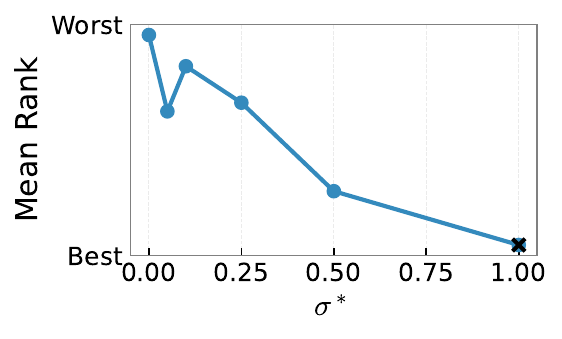}
    \caption{Supply Chain}
\end{subfigure}
\begin{subfigure}{.2\textwidth}
\includegraphics[width=\textwidth]{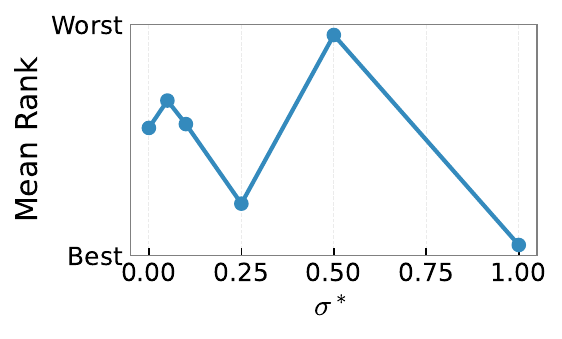}
    \caption{Duopoly}
\end{subfigure}
\begin{subfigure}{.2\textwidth}
\includegraphics[width=\textwidth]{images/calibration/triopoly_sigma.pdf}
    \caption{Triopoly}
\end{subfigure}
\begin{subfigure}{.2\textwidth}
\includegraphics[width=\textwidth]{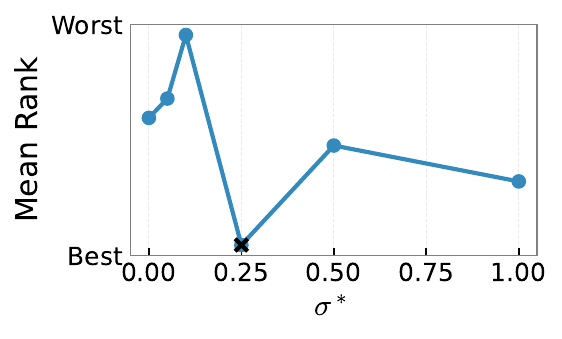}
    \caption{Cobweb}
\end{subfigure}
    \caption{Calibration Results for values of $\sigma^*$ (averaging across values of $\mu$). The lower the rank, the better.}
    \label{figCalibrationSigma}
\end{figure}

\begin{figure}
\begin{subfigure}{.2\textwidth}
    \centering
    \includegraphics[width=\textwidth]{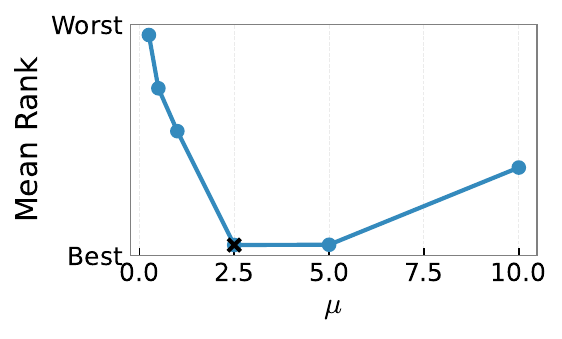}
    \caption{Supply Chain}
\end{subfigure}
\begin{subfigure}{.2\textwidth}
\includegraphics[width=\textwidth]{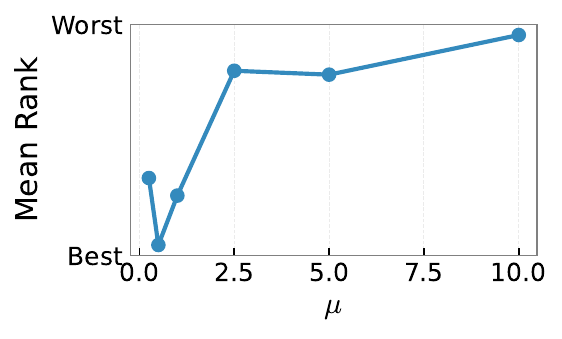}
    \caption{Duopoly}
\end{subfigure}
\begin{subfigure}{.2\textwidth}
\includegraphics[width=\textwidth]{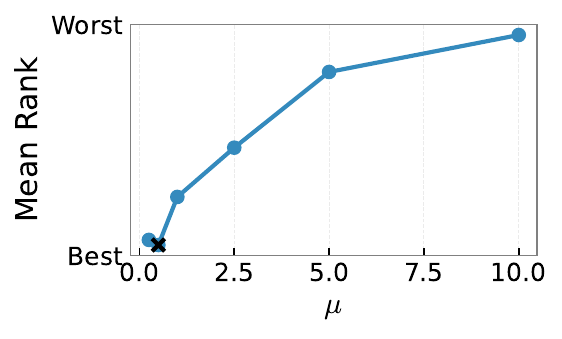}
    \caption{Triopoly}
\end{subfigure}
\begin{subfigure}{.2\textwidth}
\includegraphics[width=\textwidth]{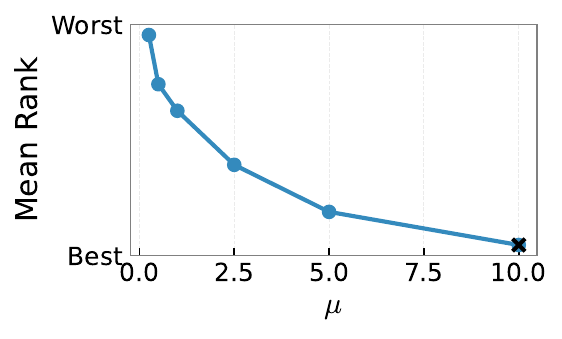}
    \caption{Cobweb}
\end{subfigure}
    \caption{Calibration Results for values of $\mu$ (averaging across values of $\sigma*$). The lower the rank, the better.}
    \label{figCalibrationMu}
\end{figure}

We visualise the results of the calibration in \cref{figCalibrationResults}. To analyse the impact of each parameter individually, we present the results for a fixed value while averaging across the other parameter range in \cref{figCalibrationSigma,figCalibrationMu}.

\end{document}